\documentclass[journal=accacs, manuscript=article, layout=twocolumn]{achemso}

\usepackage[T1]{fontenc} 
\usepackage{amsmath}
\usepackage{amssymb}
\usepackage[retain-unity-mantissa=false]{siunitx}
\usepackage{graphicx}
\usepackage{xcolor}
\usepackage[caption=false]{subfig}
\usepackage{placeins}
\usepackage[colorlinks=true, linkcolor=blue, citecolor=blue, filecolor=blue, urlcolor=blue,pdfproducer={.},pdfcreator={.}]{hyperref}
\usepackage[noabbrev, capitalize]{cleveref}
\usepackage{blindtext}
\usepackage{comment}
\usepackage{tikz}
\usepackage{verbatim}
\usepackage{ulem}
\usepackage{physics}

\newcommand{\brcom}[1]{} 

\newcommand{\ps}{\pico\second}

\renewcommand{\eqref}[1]{equation (\ref{#1})}

\xdefinecolor{rot}{RGB}{205,9,32}
\xdefinecolor{grassgreen}{RGB}{0,180,0}

\author{Christopher Seibel}
\author{Markus Uehlein}
\author{Tobias Held}
\author{Pavel N. Terekhin\footnote{Now at MCA Engineering
GmbH, 80807 Munich, Germany}}
\author{\\Sebastian T. Weber}
\author{Baerbel Rethfeld}
\email{rethfeld@rptu.de}
\affiliation[RPTU Kaiserslautern-Landau]
{Department of Physics and Research Center OPTIMAS, RPTU Kaiserslautern-Landau, 67633 Kaiserslautern, Germany}

\title{Time-resolved spectral densities of non-thermal electrons in gold}

\abbreviations{DOS}
\keywords{non-thermal electrons, hot electrons, femtosecond laser excitation of gold, energy-resolved electron dynamics,  Boltzmann collision integrals}

\begin{document}

%
%
%
%

\begin{abstract}
Noble-metal nanoparticles for photocatalysis have become a major research object in recent years due to their plasmon-enhanced strong light-matter interaction. The dynamics of the hot electrons in the noble metal are crucial for the efficiency of the photocatalysis and for the selective control of reactions. 
In this work, we present a kinetic description of the non-equilibrium electron distribution created by photoexcitation, based on full energy-resolved Boltzmann collision integrals for the laser excitation as well as for the electron-electron thermalization. 
The laser-induced electronic non-equilibrium and the inherently included secondary electron generation 
govern the dynamics of non-thermal electrons.
Applying our method to gold, we show a significant dependence of hot electron dynamics on kinetic energy. 
Specifically, the timescales of the relaxation as well as the qualitative behavior are
depending on the evaluated energy window.
During the thermalization processes there are cases of increasing electron density as well as of decreasing electron density. 
Studying the influence of excitation parameters, we find 
that the photon energy and the fluence of the exciting laser can be tuned 
to influence not only the initial excitation but also the subsequent characteristics of the time-resolved electronic spectral density dynamics. 
The electronic thermalization including secondary electron generation leads to time-dependent spectral densities which differ from their specific final equilibrium values for picoseconds after irradiation ended. 
\end{abstract}


\section{Introduction \label{sec:introdcution}}

In the face of the climate crisis, new and efficient sustainable energy sources are required to ensure a clean energy supply for future generations. In recent years, photocatalytic reactions have attracted increasing attention as a means to address this global challenge through efficient solar energy conversion~\cite{Luo2021,Zhang2019}. 
In contrast to conventional semiconductor-based photocatalysts with low photo-harvesting, noble-metal nanoparticles are promising candidates for highly efficient photocatalysts due to their strong light-matter interaction in the entire visible range~\cite{Sousa2016,Wu2015,Aslam2018}. 
In addition to solar energy harvesting, laser excitation is also used to stimulate catalytic reactions\cite{Forcherio2018,Wilson2019}.
In both cases, the light-matter interaction is enhanced by surface plasmons, providing an increased amount of photoexcited carriers for hot electron-driven catalysis~\cite{Manjavacas2014}. 
Experimentally it has been shown that the excitation conditions, e.g. the photon energy, strongly influence the hot carrier generation\cite{Minutella2017}.
If these hot carriers possess sufficient energy, they can be transferred to the unoccupied orbitals of adsorbed reactants to stimulate reactions, usually referred to as indirect carrier transfer\cite{Song2019,Zhang2018}. 
A major challenge is the short lifetime of the hot carriers
due to the fast thermalization by electron-electron scattering on femtosecond timescales\cite{Fann1992a,Brongersma2015, Bauer2015}. 
Thus, an ultrafast charge separation is required to suppress the electron-hole recombination, which can be achieved by a transfer of electrons into the conduction band of a semiconductor at a metal/semiconductor interface\cite{Li2014,Wu2015,Saunders2006}. 
To that end, the electrons must overcome the Schottky-barrier at the interface. After that they will be trapped in the conduction band resulting in an increased lifetime\cite{Sjakste2018, Park2015}. 
Hot carrier injection competes with the ultrafast decay of hot electrons in the metal and must therefore take place on even faster timescales.
Additionally, high kinetic energies are desirable for catalysis as they promote an efficient energy transfer due to fast ballistic and diffusive transport\cite{Lisowski2004}.

From all of this it is clear that future progress in photocatalysis will require a precise understanding of the excitation and thermalization processes of photoexcited electrons and in particular their timescales. To that end the temporal evolution of the energetical distribution of carriers has to be studied, in order to ultimately control the carrier dynamics and selectively steer reaction pathways.

From a theoretical point of view, the processes occurring during hot electron-driven photocatalysis are often described with  (non-adiabatic) molecular dynamics simulations\cite{Chu2020,CrespoOtero2018}, density functional theory\cite{Govorov2015,Kumar2017,DouglasGallardo2021} as well as various combinations thereof\cite{Ge2018,Aizpurua2019,Wang2021}.
However, these mostly concentrate on the catalytic pathways and reaction mechanisms assuming the existence of hot electrons without focusing on their generation and relaxation in the catalyst. 

Here we present a microscopic theory based on full Boltzmann collision integrals to describe the dynamics of electrons during excitation and thermalization in noble metals. In contrast to the relaxation time approach often used in transport calculations\cite{Singh2020,Nenno2016}, full Boltzmann integrals are able to explicitly account for the electron-electron scattering which is responsible for the thermalization.
Our model has already been successfully applied for the energy- and time-resolved kinetic description of interacting ultrafast relaxation processes in laser-excited solids for many years~\cite{Rethfeld2002, Mueller2013PRB, Weber2019}. 
By tracing the dynamics of the electronic distribution, the model allows us to determine the timescales of various processes such as for instance electron-electron and electron-phonon relaxation and their mutual influence on each other. 
Here, we apply our model to the electron-electron thermalization in gold, which is a commonly used catalyst \cite{Ge2018,Wu2015,Linic2015}.
In addition, we extract the carrier dynamics in specific energy windows to determine the time-resolved spectral densities. We find strong differences in the qualitative behavior of the temporal evolution depending on the kinetic energy.
Furthermore, we study the effect of various excitation conditions on the energy-resolved carrier dynamics, providing insights that enable custom tuning of the dynamics of laser-excited charge carriers.

\section{Model\label{sec:model}}
 During ultrashort excitation of a metal with an optical laser pulse, the electrons absorb the photons from the laser light. Thereby, the initial Fermi distribution of the electrons is perturbed, resulting in a state far from equilibrium\cite{Rethfeld2002}. 
 Through collisions with other electrons and phonons, the electrons thermalize to a Fermi distribution at elevated temperature.
 On a longer timescale of picoseconds, electrons and phonons relax to a joint thermal equilibrium\cite{Pietanza2004, Rethfeld2017}.
 The thermalization of the electrons happens on a much faster timescale than the energy transfer to the phonons\cite{Rethfeld2017,Anisimov1974}.
 Here, we are interested in the kinetic stage of the electrons focusing on their excitation and thermalization.
 Since electron-phonon coupling is weak in gold\cite{Lin2008,Mo2018}, we neglect this interaction for the following considerations.
 
The dynamics of the energy- and time-dependent electron distribution $f(E, t)$ can be described by the Boltzmann equation\cite{Rethfeld2002,Mueller2013PRB, DelFatti2000}.
We consider a homogeneously heated system, assume an isotropic momentum space and neglect external forces.
In that case, the Boltzmann equation reduces to the collision term, i.e. we calculate
\begin{equation}
    \dv{f(E, t)}{t} =  \left.\pdv{f}{t}\right|_\text{absorb} + \left.\pdv{f}{t}\right|_\text{el-el}\,, \label{eq:Boltzmann}
\end{equation}
  with full Boltzmann collision integrals for the absorption and electron-electron collision, respectively. 
  We describe the distinct collision terms below, thereby briefly summarizing the approach developed in Ref.~\citenum{Mueller2013PRB}.

\subsection{Electron-electron collisions}
There are four momenta and energies involved in the collision of two electrons. 
As in any two-particle collision, the conservation of momentum and energy is given by
\begin{align}
    \vec k + \vec k_2  &= \vec k_1 + \vec k_3 \,, \label{eq:momentum_conservation}\\
    E + E_2 &= E_1 + E_3 \,, \label{eq:energy_conservation}
\end{align}
where the incoming momenta are denoted as 
$\vec k$,~$\vec k_2$ and the outgoing momenta 
as $\vec k_1$,~$\vec k_3$, or vice versa.

Each $\vec k_i$ corresponds to an energy $E_i$. Due to the assumption of isotropy only the absolute momenta $k_i = |\vec{k_i}|$ are taken into account, which are related to the energies by the dispersion relation $E(k)$ obtained from the effective one-band model. It constructs the dispersion relation such that it reproduces the applied electronic density of states\cite{Mueller2013PRB}.

Then, the change of the electron distribution function at energy $E$, due to the collisions with other electrons, i.e., the electron-electron collision integral, is given by\cite{Mueller2013PRB}
\begin{align}
    \left.\pdv{f(E)}{t}\right|_\text{el-el} &= \frac{\Omega^2 \pi^3}{\hbar k} \int \text{d}E_1 \!\int \text{d}E_3
    \int \text{d}\Delta k \nonumber\\
    &\times \frac{D(E_1)}{k_1}\frac{D(E_2)}{k_2}\frac{D(E_3)}{k_3} \label{eq:electron_electron_collision}\\
    &\times \abs{M(\Delta k, \kappa)}^2\,\mathcal{F}\,\Xi_\text{el-el}\,, \nonumber
\end{align}
where $\Omega$ is the volume of the unit cell and $D(E)$ is the electronic density of states, which can be determined by density functional theory~(DFT)\cite{Lin2008}. The energy $E_2$ is determined by the energy conservation \cref{eq:energy_conservation}. 
The possible values for the exchanged momentum $\Delta k = |\vec{k}_1 -  \vec{k}_2 | = |\vec{k}_3 - \vec{k}|$ are limited by the step function~$\Xi_\text{el-el}$, which ensures the momentum conservation demanded by \cref{eq:momentum_conservation}\cite{Mueller2013PRB}.

The collision functional
\begin{align}
\begin{split}
    \mathcal{F} &= f_1 f_3 (1-f)(1-f_2) \\
    &- f f_2 (1-f_1) (1-f_3) \,,
\end{split}
\end{align}
determines the probability for a transition according to free and accessible states based on the Pauli exclusion principle.
Here, the $f_i$ are the electron distributions at the respective energies $E_i$ of the involved particles.
The first part of the functional describes a scattering into the considered state $E$, whereas the second part describes the scattering out of the considered state. In the latter case, the roles of incoming and outgoing particles are interchanged. 

The quantum mechanical transition probability based on the overlap of wave functions is given by the transition matrix element\cite{Mueller2013PRB}
\begin{equation}
    \abs{M(\Delta k, \kappa)}^2 = \left(\frac{e^2}{\epsilon_0 \Omega} \frac{1}{\Delta k^2 + \kappa^2}\right)\, ,
    \label{eq:matrix_element}
\end{equation}
which is derived by using a screened Coulomb potential and a plane wave approach.
It depends only on the exchanged momentum $\Delta k$ and the screening parameter $\kappa$, 
which is calculated from the given non-equilibrium distribution function\cite{DelFatti2000}.

\subsection{Absorption of photons}
We consider the absorption of photons via inverse bremsstrahlung\cite{Rethfeld2002, Seely1973, Firouzi2020}. 
In this process an electron can absorb $\ell\in\mathbb{Z}$ photons during a collision with an ion. Thus, its energy increases according to
\begin{equation}
    E_1 = E + \ell\hbar\omega\enspace,
\end{equation}
where $\omega$ is the excitation frequency. 
Note that also photoemission with $\ell<0$ is taken into account.
The energy $E_1$ corresponds to a momentum $k_1$ via the dispersion relation. The difference in momentum $\Delta k$ to the initial state is supplied by the ion, which is assumed to be located at rigid sites, thus acting 
as a source of arbitrary momentum\cite{Rethfeld2002}. 
The momentum of the photon can be considered as small and is neglected here.

For the absorption process, the collision integral reads\cite{Mueller2013PRB}
\begin{align}
    \left.\pdv{f(E)}{t}\right|_\text{absorb} &= \frac{\Omega\pi}{\hbar k} \sum_\ell \frac{D(E_1)}{k_1}\, \mathcal{F}\nonumber\\
    &\times \int \text{d}\Delta k\,\Delta k\,\abs{M(\Delta k, \kappa)}^2 \label{eq:absorption_collision}\\
    &\times \bar J_\ell^2(\gamma \Delta k)\,\Xi_\text{el-ion-photon}\,.\nonumber
\end{align}
The collision functional
\begin{equation}
    \mathcal{F} = f_1 (1-f) - f (1-f_1)
\end{equation}
considers absorption and spontaneous emission.
The probability for the absorption or emission of $\abs{\ell}$ photons is determined by the Bessel function $J_\ell$ of order $\ell$, averaged over all angles $\eta = \text{cos}( \angle(\vec{q}, \vec{E_\text{L}}))$ between the laser light field~$\vec E_\text{L}$ and the exchanged momentum $\vec q$.
It is given by
\begin{equation}
    \bar J_\ell^2(\gamma q) = \frac{1}{2} \int_{-1}^1 \text{d}\eta J_\ell^2(\gamma q \eta)
\end{equation}
with the parameter $\gamma = \frac{e \abs{\vec E_\text{L}}}{m^* \omega^2}$, which depends on the effective electron mass~$m^*$ of the given band
\cite{Rethfeld2002, Epshtein1970}.
The transition matrix element $M$ in \cref{eq:absorption_collision} is the same as for the electron-electron collision given in \cref{eq:matrix_element}\cite{Mueller2013PRB}.

The absorption strength within the material
depends on the photon energy of the exciting laser. There are two effects that are responsible for this behavior. 
Firstly, the argument $\gamma$ of the Bessel function depends on the laser frequency, which results in an increased absorption for smaller photon energies\cite{Epshtein1970}. 
Secondly, the photon energy determines the energy $E_1$ and for strong energy-dependent density of states, the term $D(E_1)$ in \cref{eq:absorption_collision} can have a large influence.
\section{Results \label{sec:results}}

With the presented model, we determine the ultrafast dynamics of gold during and after excitation with light. For the excitation, we assume an ultrashort Gaussian laser pulse, centered at $t=\SI{0}{\fs}$, with a full width at half maximum (FWHM) of \SI{50}{\fs}, as similarly used to study the dynamics of hot electrons and plasmons in photocatalysts\cite{Aguirregabiria2019,Minutella2017}.
We investigate two photon energies, $\hbar\omega_1=\SI{1.55}{\electronvolt}$ ($\mathrel{\widehat{=}} \SI{800}{\nm}$) and $\hbar\omega_2=\SI{3.1}{\electronvolt}$ ($\mathrel{\widehat{=}} \SI{400}{\nm}$), which cover the limits of the visible range. Similar ones are also used in photocatalytic experiments.\cite{Forcherio2018,Minutella2017} 
We apply fluences in the range of $F_0 = \SI{0.1}{\milli\joule\per\centi\meter\squared}$, 
which is realistic for applications in catalysis. 
This is the fluence that enters the material after reflection.
Note that the absorbed energy per volume and thus the temperature of the system after thermalization still depends on the optical properties of the material.

We calculate the changes of the electron distribution due to photon absorption and electron-electron thermalization.

\begin{figure*}
    \centering
        \includegraphics[width=0.49\textwidth]{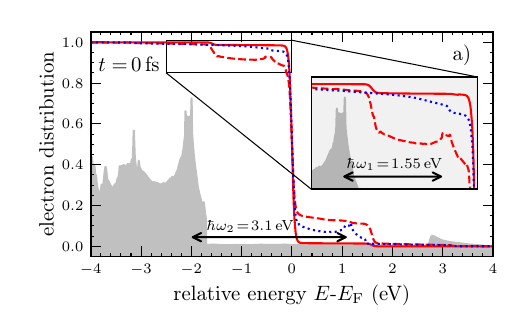}
        \includegraphics[width=0.49\textwidth]{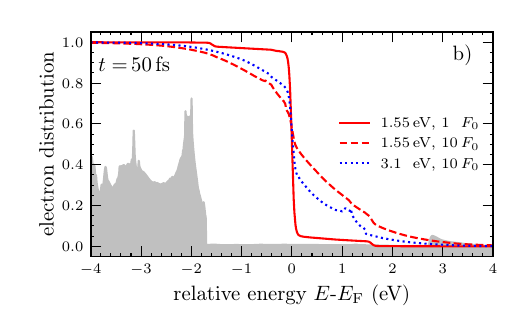}
        \includegraphics[width=0.49\textwidth]{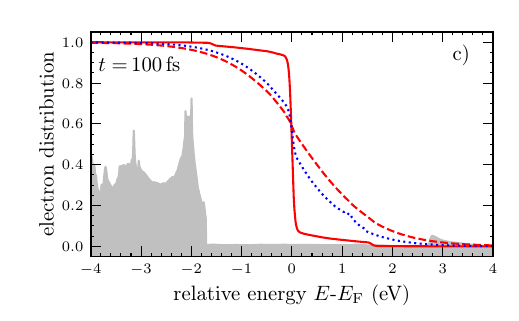}
        \includegraphics[width=0.49\textwidth]{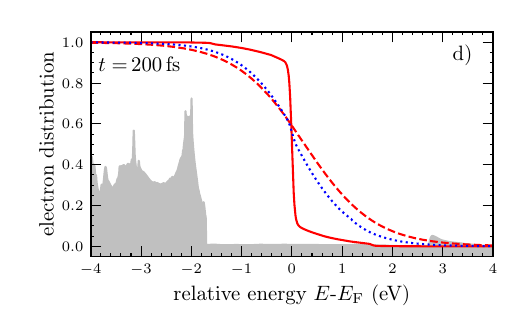}
    \caption{The electronic distribution at four instances of time a)-d) for excitations with a Gaussian laser pulse centered at $t=\SI{0}{\fs}$, with a FWHM of \SI{50}{\fs} and photon energies of $\hbar\omega_1=\SI{1.55}{\electronvolt}$ (red lines) and $\hbar\omega_2=\SI{3.1}{\electronvolt}$ (blue lines) and fluences of \mbox{$F_0 =\SI{0.1}{\milli\joule\per\centi\meter\squared}$}  (solid lines) and $10\,F_0$ (dashed and dotted lines). The used density of states\cite{Lin2008} is sketched in gray. The higher photon energy allows to depopulate the d-bands, resulting in peaks above the Fermi edge visible in panel a) at $t=\SI{0}{\fs}$. Panel b), c) and d) show the distributions at $t=\SI{50}{\fs}$, $t=\SI{100}{\fs}$ and $t=\SI{200}{\fs}$, respectively, after the pulse maximum. Their sequence reveals the influence of different excitation parameters on the thermalization process towards a respective Fermi distribution. After an excitation with a higher fluence, the thermalization is faster\cite{Mueller2013PRB}.}
    \label{fig:distributions}
\end{figure*}

\subsection{Temporal evolution of the electron distribution\label{subsec:distribution}}
\Cref{fig:distributions} shows the calculated electron distributions at four instants of 
time for different excitation conditions. Together with the distribution, we show the electronic density of states~(DOS) calculated with DFT \cite{Lin2008}. 
We apply both photon energies introduced above and fluences $F_0$ as well as $10\,F_0$. 
Note that the latter 
is used only for our theoretical demonstration since in experiments it would eventually lead to damage (melting) of the gold sample.

In \cref{fig:distributions}~a), at $t = \SI{0}{\fs}$ corresponding to the maximum intensity of the Gaussian laser pulse, we observe the effects of the laser excitation. 
Due to the Pauli exclusion principle, electrons from states below the Fermi energy~$E_\text{F}$ are excited to states above it.
In a free electron gas, 
the resulting electronic energy distribution exhibits
a structure with steps whose width is equal to the photon energy $\hbar \omega$\cite{Rethfeld2002,Fann1992b}. 
Due to the flat DOS around the Fermi energy, this step-structure is also visible in \cref{fig:distributions}~a) for $\hbar \omega_1 = \SI{1.55}{\electronvolt}$.
Comparing excitations with different fluences, we see that a larger fluence leads to a higher step. 
As soon as states below the Fermi level become available, electrons from even lower states can be excited as well. 
The more initial and/or final states can be reached, the more efficient is the excitation. Therefore, features of the DOS, such as the high peak at the d-band edge, are reflected in the excited distribution, e.g. for the higher fluence, as indicated by the arrow in the inset of \cref{fig:distributions}~a). 

With the higher photon energy~$\hbar\omega_2$, the d-states can be excited directly. 
As the d-bands provide much more initial states than the sp-bands, more electrons are excited from the
d-bands to states above Fermi edge.
In an energy interval of one photon energy above the Fermi energy
where we have seen a step-like excitation for the lower photon energy, we have two main features when a higher photon energy is applied: 
a region of higher occupation with features originating from the d-bands (between $E_\text{F}$ and $\sim E_\text{F}+\SI{1.4}{\electronvolt}$) and a region of lower occupation with excitations originating from the sp-bands ($\sim \SI{1.4}{\electronvolt}$ to $\hbar\omega_2 = \SI{3.1}{\electronvolt}$ above $E_\text{F}$).
One significant imprint of the DOS in the excited distribution is indicated by an arrow in the main part of \cref{fig:distributions}~a).

\Cref{fig:distributions}~b) to d) show the evolution after the laser pulse, i.e. the electron distribution for the same excitation condition as in \cref{fig:distributions}~a), after $50$, $100$, and $\SI{200}{\fs}$, respectively.
We observe the effect of the thermalization where the non-equilibrium features smooth out and the distributions evolve towards hot Fermi distributions. 
Our calculations confirm that a higher fluence leads to a faster thermalization\cite{Mueller2013PRB}\!\,: while the case with the lower fluence still exhibits a clear step-like structure at \SI{100}{\fs} (\cref{fig:distributions}~c)), both cases of the higher fluence show already almost thermalized distributions, i.e. equilibrium Fermi distributions of elevated temperature.
After \SI{200}{\fs}, shown in \cref{fig:distributions}~d), the distributions excited with $10\,F_0$ but different photon energies are both thermalized, however, to different temperatures.
This is indicated by the different widths of the respective Fermi distribution. 
We observe that the energy absorption is more efficient for the lower photon energy~$\hbar\omega_1$. 
This shows that the effect of the increased absorption probability for smaller photon energies discussed for the collision integral of the absorption in \cref{eq:absorption_collision} dominates here over the increased amount of states, i.e. the d-states, that could be reached with the larger photon energy.

\subsection{Generation of primary and secondary electrons\label{subsec:primary_and_secondary}}
For the transfer of electrons into adjacent materials in photocatalytic processes,
the density of electrons within a certain energy interval or above a certain energetic threshold is crucial.

\begin{figure}
    \centering
    \includegraphics{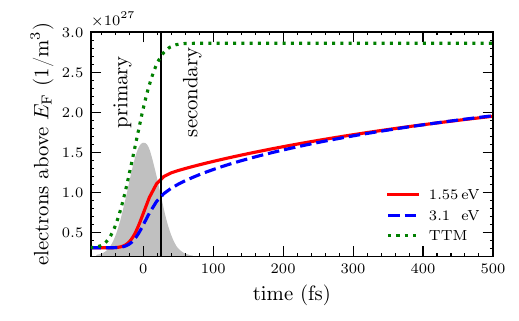}
    \caption{Evolution of electron density for all electrons above the Fermi energy for two photon energies calculated with our model as well as resulting from a two-temperature model\cite{Anisimov1974}~(TTM) for comparison. All three calculations contain the same absorbed energy.
    For the smaller photon energy $\hbar\omega_1=\SI{1.55}{\electronvolt}$, the initial generation of mainly primary electrons is more efficient whereas for the larger photon energy \mbox{$\hbar\omega_2=\SI{3.1}{\electronvolt}$} the later generation of mainly secondary electrons is more efficient, thus yielding the same density of electrons within about \SI{200}{\fs}. The TTM overestimates the number of electrons above the Fermi energy at early times due to the assumption of instantaneous relaxation.
    The shape of the laser pulse is sketched in gray.}
    \label{fig:hot_electrons}
\end{figure}

From the dynamics of the distributions, we can calculate the dynamics of the density of electrons above the Fermi energy, often called 'hot electrons'. We will refrain from using this term since it may be ambiguous: depending on the physical context it can refer either to high-energetic electrons not following an equilibrium distribution, or to electrons of a Fermi distribution at elevated temperatures. 

\Cref{fig:hot_electrons} shows the time evolution of all electrons above the Fermi energy for three different situations. In two cases (red and blue lines), we calculate the non-equilibrium dynamics with our full kinetic Boltzmann model for two respective photon energies. 
Additionally, we show the dynamics of all electrons above the Fermi energy extracted from the widely used two-temperature model (TTM) \cite{Anisimov1974}, which relies only on a temperature to describe the electronic system, i.e. it assumes an instantaneous thermalization.
Since we neglect the phonons, the TTM actually describes only a time-dependent increase of the electronic temperature. This temperature is equivalent to the 'corresponding temperature' or 'generalized temperature' applied for comparison between equilibrium and non-equilibrium distributions in, e.g., \mbox{Refs.~\citenum{Mueller2013PRB,Mueller2014ASS, Seifert2018}}.

For the calculation with the Boltzmann model and a photon energy of $\hbar\omega_1=\SI{1.55}{\electronvolt}$, we used a fluence of $F_0$.
However, applying the same fluence with a different photon energy leads to a different energy content, as seen in \cref{fig:distributions}.
To improve the comparison between the three calculations, the fluence for the non-equilibrium calculation with a photon energy of $\hbar\omega_2 =\SI{3.1}{\electronvolt}$ is adjusted to result in the same absorbed energy. 
Also the source term of the TTM has been adjusted accordingly.

During the laser pulse, electrons absorb photons. As soon as they are primarily excited, they start to generate secondary electrons, i.e. they scatter inelastically with low-energy electrons exciting them above the Fermi edge. As long as the laser pulse is active, both processes overlap but the primary electron generation is predominant. After the laser pulse, secondary electron generation holds on during the thermalization.
In \cref{fig:hot_electrons}, both electron densities calculated by the full kinetic Boltzmann model increase rapidly during the laser pulse. At approximately half of the falling edge of the pulse, a transition to a slower increase occurs, indicated by the vertical black line. 
For simplicity, we call all electrons generated before the transition "primary electron" and the ones generated after the transition "secondary electrons".

\Cref{fig:hot_electrons} shows that the generation of primary electrons is more efficient for the lower photon energy $\hbar\omega_1$, leading to more electrons above the Fermi energy at early times. 
The generation of primary electrons is less efficient for the higher photon energy $\hbar\omega_2$, because on average each electron gains more energy and thus fewer can be excited since the total absorbed energy is adjusted to be the same for both excitations.
On the other hand, the generation of secondary electrons is much more efficient for this higher photon energy.
This is because the initially high energetic electrons can more often inelastically scatter with low-energy electrons, thereby exciting more electrons above the Fermi level. Additionally, these newly excited secondary electrons can now scatter with low-energy electrons themselves, leading to an avalanche of secondary electron generation. 
The different efficiencies of primary and secondary electron generation quickly balance each other out, so that the differences in the evolution of the electron density disappear within about \SI{200}{\fs}. Beyond this time, the information of the exciting photon energy cannot be retrieved anymore from the density-evolution of the electrons above the Fermi energy. 
However, the electronic systems are not yet in equilibrium, i.e. not yet Fermi-distributed. On the one hand, this can be directly seen in \cref{fig:distributions}~d) in the red solid line, and on the other hand it is visible in \cref{fig:hot_electrons} in the difference of the electron densites to the one extracted from the TTM, which describes a fully thermalized system at all times. 

With this basic assumption, the TTM bypasses the thermalization process with the secondary electron generation or at least assumes it to be infinitely fast. 
This leads to two major shortcomings in the dynamics of the electron densities. First, the density of electrons starts to increase much too early, with the onset of the laser pulse. In the Boltzmann model, there are only a few primary electrons with high kinetic energy at this time.
Second, and more severely, the instantaneous relaxation assumed in the TTM overestimates the number of electrons above the Fermi energy in the first few hundred femtoseconds to picoseconds. 
Our calculations indicate that differences in the electron density during the thermalization process can persist for more than a picosecond.

\begin{figure}
    \centering
    \includegraphics{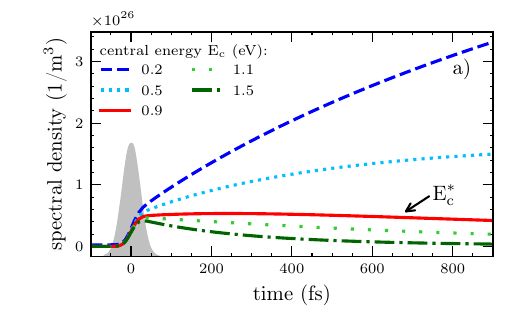}
    \includegraphics{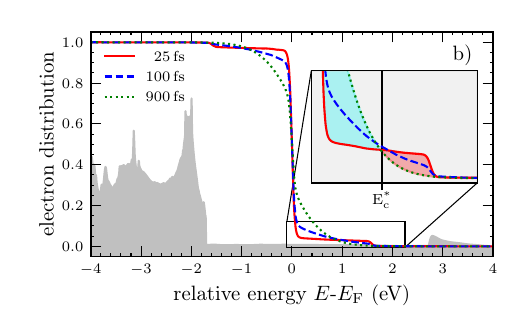}
    \caption{a) Dynamics of spectral densities above the Fermi energy after excitation with photon energy $\hbar\omega_1$ and fluence $F_0$. The central energies are given relative to the Fermi energy. A transition energy $E_c^*$ can be identified where the transition between increasing and decreasing population occurs after the laser pulse. The temporal shape of the laser pulse is sketched in gray.
    b)~Electronic distributions at various instances of time. Their intersection defines the transition energy $E_c^*$.
    Filled areas indicate spectral underpopulation (cyan) and spectral overpopulation (red) of the initially excited distribution at \SI{25}{\fs} as compared to the nearly thermalized distribution at \SI{900}{\fs}.
    The density of states is sketched in gray.}
    \label{fig:vary_spec}
\end{figure}

\subsection{Spectral electron densities \label{subsec:spec_dyn}}

In many cases, the temporal evolution of density of electrons within a certain energy window is of particalur interest. 
Therefore, we will focus here on such spectral electron densities, which can be determined from the time- and energy-dependent distribution function as
\begin{align}
    n_{E_\text{c}, \Delta E}(t) = \int\displaylimits_{E_\text{c}-\Delta E/2}^{E_\text{c}+\Delta E/2} f(E, t)\,D(E)\,{\rm d}E\enspace,
\end{align}
i.e., the density of electrons in an energy interval of width $\Delta E$ centered around a central energy $E_\text{c}$.

The resulting dynamics of the spectral densities for an excitation with photon energy $\hbar\omega_1$ are depicted in \cref{fig:vary_spec}~a) for various central energies $E_\text{c}$, given relative to the Fermi energy, each for an energetic width of \mbox{$\Delta E = \SI{0.1}{\electronvolt}$} and fluence $F_0$. 
On the timescale of the laser pulse, all spectral densities increase rapidly to approximately the same value. 
After the laser pulse, we observe a transition energy $E_c^*$ of about \SI{0.9}{\electronvolt}, where the spectral density is almost constant, exhibiting only a slight curvature (solid red line).
For energies smaller than this transition energy, i.e. closer to the Fermi edge, the densities continue to increase after the laser pulse but at a slower rate than during the pulse. 
The smallest depicted energy $E_c=\SI{0.2}{\electronvolt}$ shows the fastest and largest increase. For energies larger than the transition energy, i.e. farther from the Fermi edge, the spectral densities decrease after the laser pulse. Here, we observe a faster timescale for the largest energy $E_c = \SI{1.5}{\electronvolt}$. 
The behavior is symmetric with respect to $E_c^*$, i.e. the rate of change after the laser pulse is larger, if an energy further away from $E_c^*$ is considered. 
Note that this observation holds only in the energy range between $E_\text{F}$ and $E_\text{F}+\hbar\omega_1$, as will become clear below.

\cref{fig:vary_spec}~b) shows the non-equilibrium distribution created by the laser at different instances of time. 
At \SI{25}{\fs} (solid red line), a clear step in the distribution is visible in the inset. It originates from the excitation of electrons to the energy region mentioned above. 
The very different behavior of the spectral densities observed in \cref{fig:vary_spec}~a) results from the difference between the step created by the laser pulse and the final Fermi distribution.
At the transition energy $E_c^*$ close to \SI{0.9}{\electronvolt} (vertical black line in the inset), the distributions at different times intersect. 
This implies that the number of laser-excited electrons matches the number of electrons required for the corresponding equilibrium distribution, and thus an almost constant occupation is observed, compare also \cref{fig:vary_spec}~a). 
Below the transition energy $E_c^*$, fewer electrons are excited to a given energy state than are required for the final equilibrium distribution, thus, effectively creating a spectral underpopulation of the states compared to the final equilibrium. The cyan-filled area in the inset of \cref{fig:vary_spec}~b) illustrates this underpopulation. In contrast, at energies above the transition energy, the laser causes a spectral overpopulation (red filled area). 
The electrons in the overpopulated states scatter inelastically with lower-energetic electrons, as described above, thus leaving high-energy regions and decaying into lower-energy, underpopulated regions.
At the same time, also the secondary electrons excited from below Fermi energy gain energy and end up in the spectrally underpopulated states, thereby also contributing to the filling of these initially underpopulated states. 

In the same way, our approach also allows us to trace the dynamics of non-thermal holes below the Fermi energy, which can be used for photocatalysis similar to non-thermal electrons above the Fermi energy \cite{Tagliabue2020}.

Our observations of the spectral densities clearly show that a different relaxation time has to be considered for each energy\cite{Petek1997, Pines,Kaltenborn2014}.

\subsection{Dependence on photon energy}

In a next step, we investigate the influence of the photon energy on the temporal evolution of the spectral densities. Therefore, we compare the spectral dynamics at fixed energies but different photon energies. 

\begin{figure}
    \centering
    \includegraphics{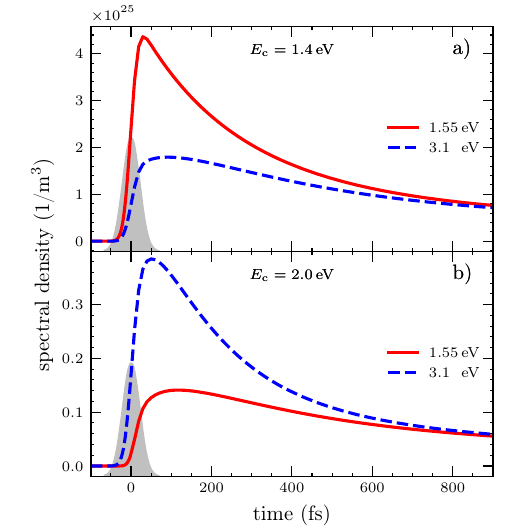}
    \includegraphics{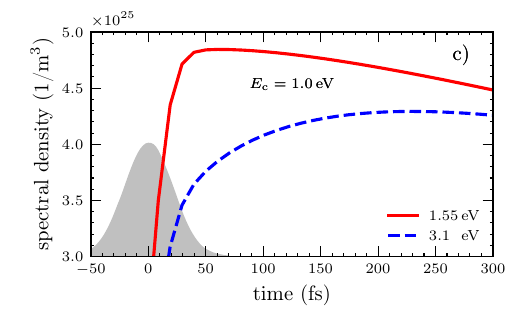}
    \caption{Spectral dynamics for two different photon energies and a fluence of $F_0$ at central energies of a) $E_\text{c}=\SI{1.4}{\electronvolt}$ and b) $E_\text{c}=\SI{2}{\electronvolt}$. In a), the smaller photon energy produces a larger spectral density whereas in b) the higher photon energy produces the larger density due to the transition from one- to two-photon absorption for the lower photon energy (see text). In c) at $E_c =\SI{1.0}{\electronvolt}$, the spectral density increases for the larger photon energy whereas it decreases for the lower one.}
    \label{fig:spectral_densities_wavelength}
\end{figure}

\Cref{fig:spectral_densities_wavelength} displays the spectral densities for excitation of gold with photon energies of \mbox{$\hbar\omega_1=\SI{1.55}{\electronvolt}$} and $\hbar\omega_2=\SI{3.1}{\electronvolt}$ at the central energies $E_\text{c}=\SI{1.4}{\electronvolt}$, $E_\text{c}=\SI{2.0}{\electronvolt}$ and $E_\text{c}=\SI{1.0}{\electronvolt}$ in a), b) and c), respectively. 
A fluence of $F_0$ was used for the \SI{1.55}{\electronvolt} simulation.
The fluence for the \SI{3.1}{\electronvolt} simulation has been adjusted to yield the same final energy content. 

In \cref{fig:spectral_densities_wavelength}~a), at the energy of \SI{1.4}{\electronvolt} above the Fermi level, the excitation with the lower photon energy creates many primary electrons, while for the larger photon energy fewer electrons are created, as similarly seen in \cref{fig:hot_electrons}. Subsequently, the generation of secondary electrons leads to a fast decay in the case of the lower photon energy. In contrast, for the case of the higher photon energy, a slower decay is observed. 
Only after almost \SI{1}{\ps}, the number of electrons is similar for both excitations, indicating that locally the differences between excitations with different photon energies can persist much longer than expected from the global picture in \cref{fig:hot_electrons}.

In \cref{fig:spectral_densities_wavelength}~b), at the energy of \SI{2.0}{\electronvolt}, a qualitatively similar behavior is observed, but in contrast to a) here the larger photon energy creates more primary electrons with a faster subsequent decay. For both photon energies, the absolute electron density is an order of magnitude lower than in the energy window evaluated in a). 

For the lower photon energy, the main difference between the two energetic positions is the transition from the one-photon to the two-photon excitation. 
For a photon energy \mbox{$\hbar\omega_1=\SI{1.55}{\electronvolt}$}, electrons can only be excited to the energy of \SI{2.0}{\electronvolt} above the Fermi energy by absorbing two photons, which is significantly less likely than a single-photon absorption. Note that the two-photon absorption is not limited here to the extremely inefficient simultaneous absorption of two photons, but also includes the consecutive absorption of one photon each at two different instances of time. 
In contrast, for the higher photon energy $\hbar\omega_2 =\SI{3.1}{\electronvolt}$ both energetic positions can be reached with a single-photon absorption. 
The difference in magnitude 
of the spectral densities excited with this photon energy to these energetic positions 
originates from the DOS of the corresponding initial states. The energy of \SI{0.4}{\electronvolt} can be reached by electrons from the d-bands, 
while the energy of \SI{2.0}{\electronvolt} can only be reached by sp-electrons. 

\cref{fig:spectral_densities_wavelength}~c) shows that we can find time- and energy regions, where the spectral density increases for the larger photon energy and decreases for the smaller one. 

\subsection{Dependence on fluence}
In addition to the photon energy, the laser fluence also plays an important role, as already seen in the temporal evolution of the excited distributions in \cref{fig:distributions}.

\begin{figure}
    \centering
        \includegraphics{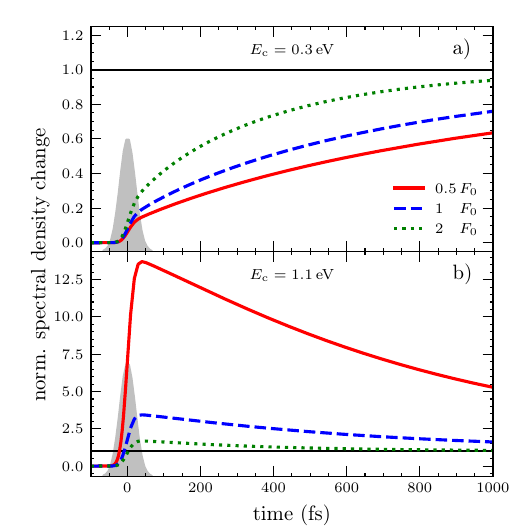}
        \includegraphics{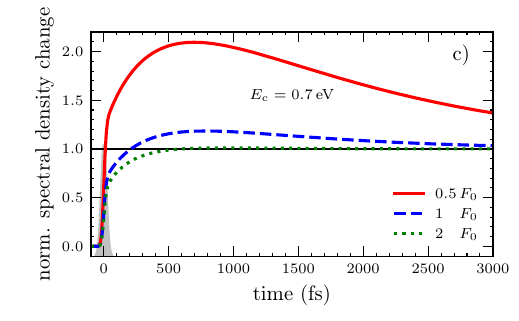}
    \caption{Spectral density change, see \cref{eq:spectral_density_change}, as normalized to its final value, for various fluences and
    a photon energy of \mbox{$\hbar\omega_1=\SI{1.55}{\electronvolt}$}. 
    We evaluate the density relaxation in different energetic regions, a) at \mbox{$E_\text{c}=\SI{0.3}{\electronvolt}$} and \mbox{b) at $E_\text{c}=\SI{1.1}{\electronvolt}$}. The results show typical traces of relaxation of initially under- and overpopulated energy regions, respectively.  
    c) In the energy region of \mbox{$E_\text{c}=\SI{0.7}{\electronvolt}$}, the initial density increase induced by the laser is followed by a slower increase due to state-filling by secondary electrons. Transiently, this leads to a spectral density exceeding the final equilibrium population (here 1.0). The subsequent relaxation to equilibrium takes places on a timescale of several picoseconds.}
    \label{fig:diff_to_therm}
\end{figure}

Thus, we investigate the dependence of the evolution of the spectral electron densities on the fluence. 
We consider the difference of the spectral density to its initial value
\begin{align}
    \Delta n_{E_\text{c}, \Delta E}(t) = n_{E_\text{c}, \Delta E}(t) - n^\text{initial}_{E_\text{c}, \Delta E} \label{eq:spectral_density_change}
\end{align}
and show this difference normalized to its final equilibrium value in \cref{fig:diff_to_therm}.
We compare the dynamics for photon energy $\hbar\omega_1=\SI{1.55}{\electronvolt}$ at three different fluences.
The normalization allows a direct comparison of the timescales of the relaxation to the different final equilibrium spectral electron densities. 
We show the spectral dynamics at three different energetic positions, which exhibit very different behavior.

\cref{fig:diff_to_therm}~a) depicts the dynamics of the spectral densities at central energy of $E_\text{c}=\SI{0.3}{\electronvolt}$.
This energy is below the transition energy $E_c^*$ identified in \cref{fig:vary_spec}~a), so a spectral underpopulation is present after the laser pulse. 
Subsequently, the secondary electron generation leads to an increase of the densities towards their final value (here unity, marked as solid black line). 
Clearly, the largest fluence results in the fastest increase, i.e. the fastest relaxation.

In \cref{fig:diff_to_therm}~b), the dynamics are shown at a central energy of $E_\text{c}=\SI{1.1}{\electronvolt}$.
This energy is above the transition energy $E_c^*$, so the laser creates a spectral overpopulation which subsequently decreases for all fluences. 
Here, the normalized overpopulation compared to the final value is largest for the smallest fluence, which decays only slowly. 

In \cref{fig:diff_to_therm}~c), at $E_\text{c}=\SI{0.7}{\electronvolt}$ close to the transition energy, we find a very different qualitative behavior, which additionally depends on the fluence. 
In all cases, the laser is responsible for a rapid increase of the spectral densities. 
At the end of the laser pulse, this results in spectral underpopulations for the larger two fluences, but in a spectral overpopulation for the lowest fluence. 
Regardless of this already-existing overpopulation, the spectral density keeps increasing further for the lowest fluence, but at a slower rate. This means that the density moves further away from the final equilibrium instead of relaxing towards it. 
For the larger two fluences, a similar increase is observed, 
which does not only approach the final equilibrium density, but exceeds it. 
The increase leading to a transient spectral overpopulation is followed by a decay towards the final equilibrium on a very long timescale of a few picoseconds.
Although the electron thermalization is usually considered to take place on a timescale of a few hundred femtoseconds, here we report traces of a non-equilibrium electron population in certain energy ranges for several picoseconds.

The reason for this behavior is the avalanche of the secondary electron generation. 
When the laser is no longer present, two competing processes determine the temporal evolution of the spectral density. First, electrons in the considered energy range can scatter to lower energies. 
Second, electrons can be gained, either when higher-energetic electrons lose energy in scattering events and decay into the considered energy range, or when lower-energetic electrons gain energy in such scattering events and are thereby excited into the considered energy range. 
At the energetic position considered in \cref{fig:diff_to_therm}~c), loss and gain of electrons are of similar importance. The increase of the population after the laser pulse beyond the equilibrium state clearly shows that the latter process dominates for several hundred femtoseconds. 
Only after \SI{0.5}{\ps} to \SI{1}{\ps}, depending on the fluence, the electron loss starts to dominate and the spectral density decreases. 
At this point, however, the entire distribution is already very close to equilibrium, so that the energy redistribution is very inefficient and therefore takes very long. 
Note that, in the presence of cold phonons, the relaxation will be faster due to an increased scattering rate and persisting cooling of the whole electron distribution. 

\section{Summary and Conclusion\label{sec:summary}}
We have presented a microscopic theory based on full energy- and time-resolved  Boltzmann collision integrals to describe the excitation and thermalization processes in metals. It is capable of tracing the full dynamics of the electronic distribution, from which the dynamics of spectral electron densities can be determined. 
Our model provides insight into the processes that determine the timescales of excitation and subsequent relaxation in gold. These timescales are particularly relevant for efficient charge separation, when carrier injection must be faster than the thermalization, i.e. the loss of high-energy electrons in a given time window. 

We have shown that the excitation conditions have a strong influence on the dynamics, a finding which is in agreement with recent experiments\cite{Minutella2017,Tagliabue2020}. 
We have found that a higher photon energy entails a less efficient generation of primary electrons, thereby initially delaying the amount of excited carriers above the Fermi energy. This is balanced out rapidly by a more efficient generation of secondary electrons. 
The TTM as a prominent example of a phenomenological model may be adequate to describe the energy balance of the whole system, especially when it comes to electron-phonon equilibration on longer timescales, but it significantly overestimates the number of excited carriers on short timescale. Note that there are already extensions of the TTM to mitigate this shortcoming by explicitly considering non-thermal electrons\cite{Carpene2006, Tsibidis2018, Uehlein2022}.
Their applicability to determine time-resolved spectral densities has to be studied in future.

Furthermore, we have shown that the dynamics of spectral electron densities exhibit strong qualitative and quantitative differences, depending on the evaluated energy range. 
They also differ from the global density dynamics considering all excited electrons. 
Our observations of these spectral densities have convincingly shown that the relaxation time in a given energetic region strongly depends on the respective energy. 
In addition, we have found energetic regions where competing processes after the excitation lead to an increase of the spectral densities several hundreds of femtoseconds after irradidation ended followed by a decrease to the final equilibrium on a picosecond timescale. 
Such behavior can not be described with a single relaxation time and, thus, questions the relaxation time approach in these energetic regions. 

In conclusion, our model is capable of tracing the dynamics in light-excited 
metals
by explicitly considering microscopic collision integrals.
It can be applied in the design of novel catalyst systems for an enhanced carrier injection efficiency by predicting the carrier dynamics in certain energy windows for different excitation conditions. 
This leads
to the perspective to tune these parameters in order to achieve custom electron density distributions and
dynamics.

\begin{acknowledgement}

Financial support of the Deutsche Forschungsgemeinschaft (DFG,  German Research  Foundation) 
through the SFB/TRR-173-268565370 “Spin+X” (Project No. A08 and B03) is gratefully acknowledged.
We appreciate the Allianz für
Hochleistungsrechnen Rheinland-Pfalz for providing 
computing resources through project STREMON
on the Elwetritsch high performance computing cluster.
\end{acknowledgement}

%
%
%

\bibliography{bibfile/all.bib,photocatalysis.bib}

\end{document}